\documentclass[11pt]{article}
\usepackage{amssymb}
\usepackage{amsmath}
\usepackage{mymacro}

\begin{document}
\sloppy

\title{\Large\bf
  Quantum Multi-Prover Interactive Proof Systems\\
  with Limited Prior Entanglement
}

\author{
{\large \hspace*{-1ex} $\mbox{\bf Hirotada Kobayashi}^{\ast \dagger}$
 \hspace*{-1ex}}\\
{\tt \hspace*{-2ex} hirotada@qci.jst.go.jp \hspace*{-2ex}} 
\and
{\large \hspace*{-1ex} $\mbox{\bf Keiji Matsumoto}^{\ddagger \ast}$
 \hspace*{-1ex}}\\
{\tt \hspace*{-2ex} keiji@nii.ac.jp \hspace*{-2ex}}
}

\date{}

\maketitle
\thispagestyle{plain}
\pagestyle{plain}

\begin{center}
{\large
  ${}^{\ast}$Quantum Computation and Information Project\\
  Exploratory Research for Advanced Technology\\
  Japan Science and Technology Corporation\\
  5-28-3 Hongo, Bunkyo-ku, Tokyo 113-0033, Japan
}

\vspace{5mm}
{\large
  ${}^{\dagger}$Department of Information Science\\
  Graduate School of Science\\
  The University of Tokyo\\
  7-3-1 Hongo, Bunkyo-ku, Tokyo 113-0033, Japan
}

\vspace{5mm}
{\large
  ${}^{\ddagger}$Foundations of Information Research Division\\
  National Institute of Informatics\\
  2-1-2 Hitotsubashi, Chiyoda-ku, Tokyo 101-8430, Japan
}

\vspace{8mm}
{\large 6 June 2003}
\end{center}

\vspace{3mm}

\begin{abstract}

This paper gives the first formal treatment of
a quantum analogue of multi-prover interactive proof systems.
It is proved that the class of languages
having quantum multi-prover interactive proof systems
is necessarily contained in $\NEXP$,
under the assumption that provers are allowed to share
at most polynomially many prior-entangled qubits.
This implies that, in particular,
if provers do not share any prior entanglement with each other,
the class of languages
having quantum multi-prover interactive proof systems is equal to $\NEXP$.
Related to these, it is shown that,
in the case a prover does not have his private qubits,
the class of languages
having quantum single-prover interactive proof systems
is also equal to $\NEXP$.

\end{abstract}



\section{Introduction}
\label{Section: Introduction}

After Deutsch~\cite{Deu85PRSLA} gave the first formal treatment of quantum computation,
a number of papers have provided evidence that
quantum computation has much more power than classical computation
for solving certain computational tasks,
including notable Shor's integer factoring algorithm~\cite{Sho97SIComp}.
Watrous~\cite{Wat03TCS} showed that
it might be also the case for single-prover interactive proof systems,
by constructing a constant-round quantum interactive protocol
for a $\PSPACE$-complete language,
which is impossible for classical interactive proof systems
unless the polynomial-time hierarchy collapses to $\AM$~\cite{Bab85STOC, GolSip89RC}.
A natural question to ask is how strong
a quantum analogue of multi-prover interactive proof systems is.
This paper gives the first step for this question,
by proving that the class of languages
having quantum multi-prover interactive proof systems
is necessarily contained in non-deterministic exponential time ($\NEXP$),
under the assumption that provers are allowed to share
at most polynomially many prior-entangled qubits.
This might even suggest that, under such an assumption,
quantum multi-prover interactive proof systems
are weaker than classical ones,
since Cleve~\cite{Cle00WQCI} reported that
a pair of provers sharing polynomially many entangled qubits
can in some sense cheat a classical verifier.

Interactive proof systems were introduced by Babai~\cite{Bab85STOC}
and Goldwasser, Micali, and Rackoff~\cite{GolMicRac89SIComp}.
An interactive proof system consists of an interaction
between a computationally unbounded prover and a polynomial-time probabilistic verifier.
The prover attempts to convince the verifier that
a given input string satisfies some property,
while the verifier tries to verify the validity of the assertion of the prover.
A language $L$ is said to have an interactive proof system
if there exists a verifier $V$ such that
(i) in the case the input is in $L$,
there exists a prover $P$ that can convince $V$
with certainty,
and (ii) in the case the input is not in $L$,
no prover $P'$ can convince $V$ with probability more than $1/2$.
It is well-known that
the class of languages having interactive proof systems, denoted by $\IP$,
is equal to $\PSPACE$,
shown by Shamir~\cite{Sha92JACM}
based on the work of Lund, Fortnow, Karloff, and Nisan~\cite{LunForKarNis92JACM},
and on the result of Papadimitriou~\cite{Pap85JCSS}
(see also~\cite{She92JACM}).

Quantum interactive proof systems were introduced by Watrous~\cite{Wat03TCS}
in terms of quantum circuits.
He showed that
every language in $\PSPACE$ has a quantum interactive protocol,
with exponentially small one-sided error, in which the prover and the verifier
exchange only three messages.
A consecutive work of Kitaev and Watrous~\cite{KitWat00STOC}
showed that any quantum interactive protocol,
even with two-sided bounded error,
can be parallelized to a three-message
quantum protocol with exponentially small one-sided error.
They also showed that the class of languages
having quantum interactive proof systems is necessarily contained
in deterministic exponential time ($\EXP$).

A multi-prover interactive proof system,
introduced by Ben-Or, Goldwasser, Kilian, and Wigderson~\cite{BenGolKilWig88STOC},
is an extension of a (single-prover) interactive proof system
in which a verifier communicates with not only one
but multiple provers,
while provers cannot communicate with each other prover
and cannot know messages exchanged between the verifier and other provers.
A language $L$ is said to have a multi-prover interactive proof system
if, for some $k$ denoting the number of provers,
there exists a verifier $V$ such that
(i) in the case the input is in $L$,
there exist provers $P_{1}, \ldots, P_{k}$ that can convince $V$
with certainty,
and (ii) in the case the input is not in $L$,
no set of provers $P'_{1}, \ldots, P'_{k}$ can convince $V$ with probability more than $1/2$.
Babai, Fortnow, and Lund~\cite{BabForLun91CC},
combining the result by Fortnow, Rompel, and Sipser~\cite{ForRomSip94TCS},
showed that
the class of languages
having multi-prover interactive proof systems, denoted by $\MIP$,
is equal to $\NEXP$.
A sequence of papers by Cai, Condon, and Lipton~\cite{CaiConLip90SCT, CaiConLip94JCSS},
Feige~\cite{Fei91SCT}, and Lapidot and Shamir~\cite{LapSha97JCSS}
led to a result of Feige and Lov\'{a}sz~\cite{FeiLov92STOC}
that every language in $\NEXP$ has a two-prover interactive proof system
with just one round (i.e. two messages) of communication
(meaning that the verifier sends one question to each of the provers in parallel,
then receives their responses), with exponentially small one-sided error.

In this paper we first define quantum multi-prover interactive proof systems
by naturally extending the quantum single-prover model.
Perhaps the most important and interesting difference between
quantum and classical multi-prover interactive proofs
is that provers may share entanglement {\em a priori\/}.
Particular cases are
protocols with two provers
initially sharing lots of EPR pairs.
For the sake of generality,
we may allow protocols with any number of provers
and with any kind of prior entanglement, not limited to EPR-type ones.
Although sharing classical randomness among provers
does not change the power of classical multi-prover interactive proofs
(unless zero-knowledge properties are taken into account~\cite{BelFeiKil95ISTCS}),
sharing prior entanglement does have a possibility
both to strengthen and to weaken the power of
quantum multi-prover interactive proofs.
In fact,
while sharing prior entanglement may increase the power of cheating provers
as shown by Cleve~\cite{Cle00WQCI},
it may be possible for a quantum verifier
to turn the prior entanglement among provers to his advantage.

The main result of this paper is
to show the $\NEXP$ upper bound for 
quantum multi-prover interactive proof systems
under the assumption that provers are allowed to share
at most polynomially many prior-entangled qubits.
That is, polynomially many prior-entangled qubits among provers
cannot be advantageous to a quantum verifier.
As a special case of this result,
it is proved that,
if provers do not share any prior entanglement with each other,
the class of languages
having quantum multi-prover interactive proof systems is equal to $\NEXP$.
Another result related to these is that,
in the case the prover does not have his private qubits,
the class of languages
having quantum single-prover interactive proof systems
is also equal to $\NEXP$.
This special model of quantum single-prover interactive proofs
can be regarded as a quantum counterpart
of a probabilistic oracle machine~\cite{ForRomSip94TCS, For89PhD, BabForLun91CC}
in the sense that there is no private space for the prover
during the protocol,
and thus we call this model as a {\em quantum oracle circuit\/}.
Our result shows that quantumization of probabilistic oracle machines
does not change the power of the model.

To prove the $\NEXP$ upper bound of quantum multi-prover interactive proof systems,
a key idea is to bound the number of private qubits of provers
without diminishing the computational power of them.
Suppose that each prover has only polynomially many private qubits
during the protocol.
Then the total number of qubits of the quantum multi-prover interactive proof system
is polynomially bounded,
and we can show that it can be simulated classically in non-deterministic exponential time.
Now the point is whether
space-bounded quantum provers
(i.e. provers can apply any unitary transformations on their spaces,
but the number of qubits in their spaces is bounded polynomial with respect to the input length)
are as powerful as space-unbounded quantum provers or not.
Under the assumption that provers are allowed to share
at most polynomially many prior-entangled qubits,
we show that, even with only polynomially many private qubits,
each prover can do everything that he could with as many qubits as he likes,
in the sense that the verifier cannot distinguish the difference at all.
For this, we also prove one fundamental property on quantum information theory
using the entanglement measure introduced by Nielsen~\cite{Nie01QIP}.
Apart from quantum interactive proof systems,
this property itself is also of interest and worth while stating.

The remainder of this paper is organized as follows.
In Section~\ref{Section: Quantum Fundamentals}
we briefly review basic notations and definitions
in quantum computation and quantum information theory.
In Section~\ref{Section: Definitions}
we give a formal definition of quantum multi-prover interactive proof systems and quantum oracle circuits.
In Section~\ref{Section: NEXP upper bound}
we show our main result of
the $\NEXP$ upper bound of quantum multi-prover interactive proof systems.
In Section~\ref{Section: QMIP(n.e.) = QOC = NEXP}
we focus on the prior unentangled cases and on quantum oracle circuits.
Finally we conclude with Section~\ref{Section: Conclusions},
which summarizes our results
and mentions a number of open problems related to our work.


\section{Quantum Fundamentals}
\label{Section: Quantum Fundamentals}

Here we briefly review basic notations and definitions
in quantum computation and quantum information theory.
Detailed descriptions are, for instance,
in~\cite{Gru99Book, NieChu00Book, KitSheVya02Book}.

A {\em pure state\/} is described by a unit vector in some Hilbert space.
In particular, an $n$-dimensional pure state is a unit vector $\ket{\psi}$
in $\Complex^{n}$.
Let $\{ \ket{e_{1}}, \ldots, \ket{e_{n}} \}$
be an orthonormal basis for $\Complex^{n}$.
Then any pure state in $\Complex^{n}$ can be described as
$\sum_{i=1}^{n} \alpha_{i} \ket{e_{i}}$
for some
$\alpha_{1}, \ldots, \alpha_{n} \in \Complex,
\sum_{i=1}^{n} \abs{\alpha_{i}}^{2} = 1$.

A {\em mixed state\/} is a classical probability distribution
$(p_{i}, \ket{\psi_{i}})$, $0 \leq p_{i} \leq 1$, $\sum_{i}p_{i} = 1$
over pure states $\ket{\psi_{i}}$.
This can be interpreted as being in the pure state
$\ket{\psi_{i}}$ with probability $p_{i}$.
A mixed state is often described in the form of a {\em density matrix\/}
$\rho = \sum_{i}p_{i} \ketbra{\psi_{i}}$.
Any density matrix is positive semidefinite and has trace $1$.

If a unitary transformation $U$ is applied to a state $\ket{\psi}$,
the state becomes $U \ket{\psi}$.
In the form of density matrices,
a state $\rho$ changes to $U \rho \conjugate{U}$ after $U$ is applied.

One of the important operations to density matrices is
the {\em trace-out\/} operation.
Given a density matrix $\rho$ over $\calH \otimes \calK$,
the state after tracing out $\calK$ is a density matrix over $\calH$
described by
\[
\tr_{\calK} \rho
= \sum_{i=1}^{n} (I_{\calH} \otimes \bra{e_{i}})
\rho (I_{\calH} \otimes \ket{e_{i}})
\]
for any orthonormal basis $\{ \ket{e_{1}}, \ldots, \ket{e_{n}} \}$ of $\calK$,
where $n$ is the dimension of $\calK$
and $I_{\calH}$ is the identity operator over $\calH$.
To perform this operation on some part of a quantum system
gives a partial view of the quantum system with respect to the remaining part.

One of the important concepts in quantum physics is a {\em measurement\/}.
Any collection of linear operators $\{ A_{1}, \ldots, A_{k} \}$
satisfying $\sum_{i=1}^{k} \conjugate{A_{i}} A_{i} = I$ defines a measurement.
If a system is in a pure state $\ket{\psi}$,
such a measurement results in $i$ with probability $\norm{A_{i} \ket{\psi}}^{2}$,
and the state becomes $A_{i} \ket{\psi} / \norm{A_{i} \ket{\psi}}$.
If a system is in a mixed state with a density matrix $\rho$,
the result $i$ is observed with probability $\tr(A_{i} \rho \conjugate{A_{i}})$,
and the state after the measurement is with a density matrix
$A_{i} \rho \conjugate{A_{i}} / \tr(A_{i} \rho \conjugate{A_{i}})$.
A special class of measurements are {\em projection\/} or {\em von Neumann\/} measurements
in which $\{ A_{1}, \ldots, A_{k} \}$ is a collection of orthonormal projections.
In this scheme,
an observable is a decomposition of $\calH$ into orthogonal subspaces
$\calH_{1}, \ldots, \calH_{k}$, that is,
$\calH = \calH_{1} \oplus \cdots \oplus \calH_{k}$.
It is important to note
that two mixed states having the identical density matrix
cannot be distinguished at all by any measurement. 

For any linear operator $A$ over $\calH$, the $l_{2}$-norm of $A$ is defined by
\[
\norm{A} = \sup_{\ket{\psi} \in \calH \setminus \{0\}}
\frac{\norm{A \ket{\psi}}}{\norm{\ket{\psi}}}.
\]


\section{Definitions}
\label{Section: Definitions}

\subsection{Polynomial-Time Uniformly Generated Families of Quantum Circuits}
\label{Subsection: Uniform QC}

Similar to the model of quantum single-prover interactive proof systems
discussed in~\cite{Wat03TCS, KitWat00STOC},
we define quantum multi-prover interactive proof systems
in terms of quantum circuits.
Before proceeding to the definition of quantum multi-prover interactive proof systems,
we review the concept of polynomial-time uniformly generated families of quantum circuits.

A family $\{ Q_{x} \}$ of quantum circuits is
{\em polynomial-time uniformly generated\/}
if there exists a deterministic procedure
that, on every input $x$, outputs a description of $Q_{x}$
and runs in time polynomial in $n = |x|$.
For simplicity, we assume all input strings are over the alphabet $\Sigma = \{ 0, 1 \}$.
It is assumed that the circuits in such a family are composed of gates
in some reasonable, universal, finite set of quantum gates
such as the Shor basis~\cite{Sho96FOCS, BoyMorPulRoyVat00IPL}:
Hadamard gates, $\sqrt{\sigma_{z}}$ gates, and Toffoli gates.
Furthermore, it is assumed that the number of gates in any circuit
is not more than the length of the description of that circuit.
Therefore $Q_{x}$ must have size polynomial in $n$.
For convenience,
we may identify a circuit $Q_{x}$ with the unitary operator it induces.

It should be mentioned that
to permit non-unitary quantum circuits, in particular,
to permit measurements at any timing during the computation
does not change the computational power of the model
in view of time complexity.
See~\cite{AhaKitNis98STOC} for a detailed description
of the equivalence of the unitary and non-unitary quantum circuit models.

\subsection{Quantum Multi-Prover Interactive Proof Systems}
\label{Subsection: QMIP definition}

Here we give a formal definition
of quantum multi-prover interactive proof systems
which is a natural extension of quantum single-prover ones defined by Watrous~\cite{Wat03TCS}.
In fact, the model of quantum single-prover interactive proof systems
discussed in~\cite{Wat03TCS, KitWat00STOC} is a special case of our quantum multi-prover model
with the restriction of the number of provers to one.

Let $k$ be the number of provers.
For every input $x \in \Sigma^{\ast}$ of length $n = |x|$,
the entire system of quantum $k$-prover interactive proof system consists of
$q(n) = q_{\calV}(n) + \sum_{i=1}^{k} (q_{\calM_{i}}(n) + q_{\calP_{i}}(n))$
qubits,
where $q_{\calV}(n)$ is the number of qubits
that are private to a verifier $V$,
each $q_{\calP_{i}}(n)$ is the number of qubits
that are private to a prover $P_{i}$,
and each $q_{\calM_{i}}(n)$ is the number of message qubits
used for communication between $V$ and $P_{i}$.
Note that no communication is allowed between different provers $P_{i}$ and $P_{j}$.
It is assumed that $q_{\calV}$ and each $q_{\calM_{i}}$ are polynomially bounded functions.
Moreover, without loss of generality, we may assume that
$q_{\calM_{1}} = \cdots = q_{\calM_{k}} = q_{\calM}$
and $q_{\calP_{1}} = \cdots = q_{\calP_{k}} = q_{\calP}$.
Accordingly, the entire system consists of
$q(n) = q_{\calV}(n) + k (q_{\calM}(n) + q_{\calP}(n))$
qubits.

Given polynomially bounded functions
$m, q_{\calV}, q_{\calM} \colon \Nonnegative \rightarrow \Natural$,
an {\em $m$-message $(q_{\calV}, q_{\calM})$-restricted quantum verifier\/} $V$
for a quantum $k$-prover interactive proof system
is a polynomial-time computable mapping of the form
$V \colon \Sigma^{\ast} \rightarrow \Sigma^{\ast}$,
where $\Sigma = \{ 0, 1 \}$ is the alphabet set.
For every input $x \in \Sigma^{\ast}$ of length $n$,
$V$ uses at most $q_{\calV}(n)$ qubits for his private space
and at most $q_{\calM}(n)$ qubits for communication with each prover.
The string $V(x)$ is interpreted as a $\lfloor m(n)/2 + 1 \rfloor$-tuple
$(V(x)_{1}, \ldots, V(x)_{\lfloor m(n)/2 + 1 \rfloor})$,
with each $V(x)_{j}$ a description of a polynomial-time uniformly generated
quantum circuit acting on
$q_{\calV}(n) + k q_{\calM}(n)$ qubits.
One of the private qubits of the verifier is designated as the output qubit.

Given polynomially bounded functions
$m, q_{\calM} \colon \Nonnegative \rightarrow \Natural$
and a function $q_{\calP} \colon \Nonnegative \rightarrow \Natural$,
an {\em $m$-message $(q_{\calM}, q_{\calP})$-restricted quantum prover\/} $P_{i}$
for each $i = 1, \ldots, k$ is a mapping of the form
$P_{i} \colon \Sigma^{\ast} \rightarrow \Sigma^{\ast}$.
For every input $x \in \Sigma^{\ast}$ of length $n$,
each $P_{i}$ uses at most $q_{\calP}(n)$ qubits for his private space
and at most $q_{\calM}(n)$ qubits for communication with the verifier.
The string $P_{i}(x)$ is interpreted as a $\lfloor m(n)/2 + 1/2 \rfloor$-tuple
$(P_{i}(x)_{1}, \ldots, P_{i}(x)_{\lfloor m(n)/2 + 1/2 \rfloor})$,
with each $P_{i}(x)_{j}$ a description of a quantum circuit
acting on $q_{\calM}(n) + q_{\calP}(n)$ qubits.
No restrictions are placed on the complexity of the mapping $P_{i}$
(i.e., each $P_{i}(x)_{j}$ can be an arbitrary unitary transformation).
Furthermore,
for some function
$q_{\ent} \colon \Nonnegative \rightarrow \Natural$
satisfying $q_{\ent} \leq q_{\calP}$,
each $P_{i}$ may have at most $q_{\ent}(n)$ qubits among his private qubits
that are prior-entangled with some private qubits of other provers.
Such a prover $P_{i}$ is said {\em $q_{\ent}$-prior-entangled\/}.
For the sake of generality,
we allow any kind of prior entanglement, not limited to EPR-type ones.

\begin{figure}[t]
\begin{center}
\setlength{\unitlength}{0.84mm}
{\small
\begin{picture}(163.5, 80)(-2, 0)
\put( -2,65){\makebox(30, 8){\shortstack{private qubits\\ of prover 1}}}
\put( -2,50){\makebox(30, 8){\shortstack{message qubits\\ with prover 1}}}
\put( -2,35){\makebox(30, 8){\shortstack{private qubits\\ of verifier}}}
\put( -2,20){\makebox(30, 8){\shortstack{message qubits\\ with prover 2}}}
\put( -2, 5){\makebox(30, 8){\shortstack{private qubits\\ of prover 2}}}
\multiput( 30,65)(0,2){5}{\line(1,0){5}}
\multiput( 30,50)(0,2){5}{\line(1,0){5}}
\multiput( 30,35)(0,2){5}{\line(1,0){30}}
\multiput( 30,20)(0,2){5}{\line(1,0){5}}
\multiput( 30, 5)(0,2){5}{\line(1,0){5}}
\put( 35,46.5){\framebox(15,30){$P_{1}(x)_{1}$}}
\put( 35, 1.5){\framebox(15,30){$P_{2}(x)_{1}$}}
\multiput( 50,65)(0,2){5}{\line(1,0){35}}
\multiput( 50,50)(0,2){5}{\line(1,0){10}}
\multiput( 50,20)(0,2){5}{\line(1,0){10}}
\multiput( 50, 5)(0,2){5}{\line(1,0){35}}
\put( 60,16.5){\framebox(15,45){$V(x)_{1}$}}
\multiput( 75,50)(0,2){5}{\line(1,0){10}}
\multiput( 75,35)(0,2){5}{\line(1,0){35}}
\multiput( 75,20)(0,2){5}{\line(1,0){10}}
\put( 85,46.5){\framebox(15,30){$P_{1}(x)_{2}$}}
\put( 85, 1.5){\framebox(15,30){$P_{2}(x)_{2}$}}
\multiput(100,65)(0,2){5}{\line(1,0){30}}
\multiput(100,50)(0,2){5}{\line(1,0){10}}
\multiput(100,20)(0,2){5}{\line(1,0){10}}
\multiput(100, 5)(0,2){5}{\line(1,0){30}}
\put(110,16.5){\framebox(15,45){$V(x)_{2}$}}
\multiput(125,50)(0,2){5}{\line(1,0){5}}
\multiput(125,35)(0,2){5}{\line(1,0){5}}
\multiput(125,20)(0,2){5}{\line(1,0){5}}
\put(135,48){\vector(-1,-1){4.5}}
\put(133.5,48){\makebox(28,5){output qubit}}
\end{picture}
}
\caption{Quantum circuit for a three-message quantum two-prover interactive proof system}
\label{Figure: QMIPfigure}
\end{center}
\end{figure}
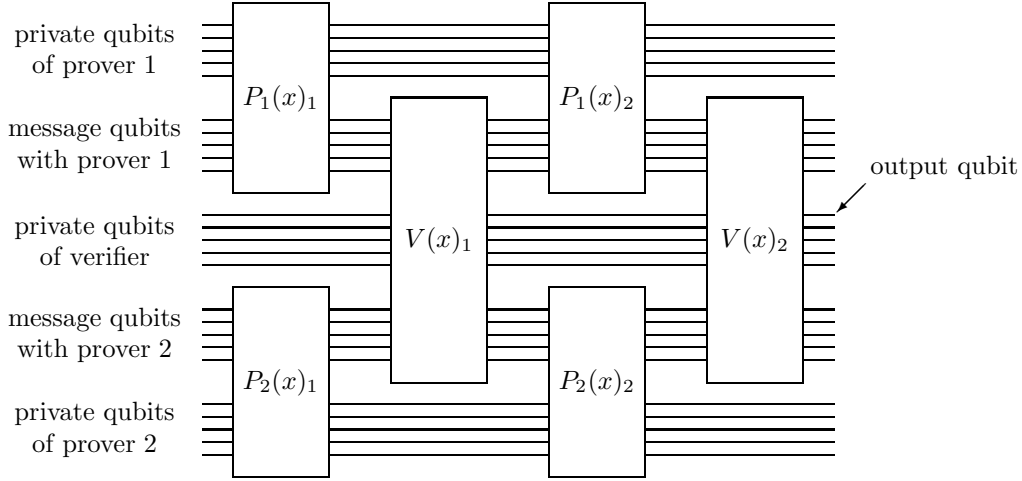

An {\em $m$-message $(q_{\calV}, q_{\calM}, q_{\calP})$-restricted
quantum $k$-prover interactive proof system\/}
consists of 
an $m$-message $(q_{\calV}, q_{\calM})$-restricted quantum verifier $V$
and $m$-message $(q_{\calM}, q_{\calP})$-restricted quantum provers
$P_{1}, \ldots, P_{k}$.
If $P_{1}, \ldots, P_{k}$ are $q_{\ent}$-prior-entangled,
such a quantum $k$-prover interactive proof system is said
{\em $q_{\ent}$-prior-entangled\/}.
Let $\calV = l_{2}(\Sigma^{q_{\calV}})$,
each $\calM_{i} = l_{2}(\Sigma^{q_{\calM}})$,
and each $\calP_{i} = l_{2}(\Sigma^{q_{\calP}})$
denote the Hilbert spaces
corresponding to the private qubits of the verifier,
the message qubits between the verifier and the $i$th prover,
and the private qubits of the $i$th prover, respectively. 
Given a verifier $V$, provers $P_{1}, \ldots, P_{k}$, and an input $x$ of length $n$,
we define a circuit $(P_{1}(x), \ldots, P_{k}(x), V(x))$ acting on $q(n)$ qubits as follows.
If $m(n)$ is odd, circuits
${P_{1}(x)_{1}, \ldots, P_{k}(x)_{1}}$, $V(x)_{1}$,
$\ldots$,
${P_{1}(x)_{(m(n)+1)/2}, \ldots, P_{k}(x)_{(m(n)+1)/2}}$,
$V(x)_{(m(n)+1)/2}$
are applied in sequence, each $P_{i}(x)_{j}$ to $\calM_{i} \otimes \calP_{i}$,
and each $V(x)_{j}$ to
$\calV \otimes {\calM_{1} \otimes \cdots \otimes \calM_{k}}$.
If $m(n)$ is even, circuits
$V(x)_{1}$,
${P_{1}(x)_{1}, \ldots, P_{k}(x)_{1}}$,
$\ldots$,
$V(x)_{m(n)/2}$,
${P_{1}(x)_{m(n)/2}, \ldots, P_{k}(x)_{m(n)/2}}$,
$V(x)_{m(n)/2 + 1}$
are applied in sequence.
Figure~\ref{Figure: QMIPfigure}
illustrates the situation for the case $k=2$ and $m(n) = 3$.
Note that the order of applications of the circuits of the provers at each round
has actually no sense
since the space $\calM_{i} \otimes \calP_{i}$
on which the circuits of the $i$th prover act is separated from each other prover.

At any given instant, the state of the entire system is
a unit vector in the space
$\calV \otimes \calM_{1} \otimes \cdots \otimes \calM_{k}
\otimes \calP_{1} \otimes \cdots \otimes \calP_{k}$.
For instance, in the case $m(n) = 3$, given an input $x$ of length $n$,
the state of the system after all of the circuits of the provers and the verifier
have been applied is
\[
V_{2} P_{k,2} \cdots P_{1,2} V_{1} P_{k,1} \cdots P_{1,1} \ket{\init},
\]
where each $V_{j}$ and $P_{i,j}$ denotes the extension of $V(x)_{j}$
and $P_{i}(x)_{j}$, respectively,
to the space
$\calV \otimes {\calM_{1} \otimes \cdots \otimes \calM_{k}}
\otimes {\calP_{1} \otimes \cdots \otimes \calP_{k}}$
by tensoring with the identity,
and $\ket{\init} \in \calV \otimes {\calM_{1} \otimes \cdots \otimes \calM_{k}}
\otimes {\calP_{1} \otimes \cdots \otimes \calP_{k}}$
denotes the initial state.
In the initial state $\ket{\init}$
for $q_{\ent}$-prior-entangled proof systems,
only the first $q_{\ent}(n)$ qubits in each $\calP_{i}$
may be entangled with other qubits in
$\calP_{1} \otimes \cdots \otimes \calP_{k}$.
All the qubits other than these prior-entangled ones
are initially in the $\ket{0}$-state.

For every input $x$, the probability
that the $(k+1)$-tuple $(P_{1}, \ldots, P_{k}, V)$ accepts $x$ is defined
to be the probability that an observation of the output qubit
in the basis of $\{ \ket{0}, \ket{1} \}$ yields $\ket{1}$,
after the circuit $(P_{1}(x), \ldots, P_{k}(x), V(x))$
is applied to the initial state $\ket{\init}$.

Although $k$, the number of provers, has been treated to be constant so far,
the above definition can be naturally extended to the case that
$k \colon \Nonnegative \rightarrow \Natural$
is a function of the input length $n$.
In what follows, we treat $k$ as a function.
Note that the number of provers possible to communicate with the verifier
must be bounded polynomial in $n$.

\begin{definition}
Given polynomially bounded functions
$k, m \colon \Nonnegative \rightarrow \Natural$,
a function $q_{\ent} \colon \Nonnegative \rightarrow \Natural$,
and functions $a, b \colon \Nonnegative \rightarrow [0,1]$,
a language $L$ is in $\QMIP(k, m, q_{\ent}, a, b)$
iff there exist polynomially bounded functions
$q_{\calV}, q_{\calM} \colon \Nonnegative \rightarrow \Natural$
and an $m$-message $(q_{\calV}, q_{\calM})$-restricted quantum verifier $V$
for a quantum $k$-prover interactive proof system
such that, for every input $x$ of length $n$,
\begin{itemize}
\item[(i)]
if $x \in L$, there exist a function
$q_{\calP} \colon \Nonnegative \rightarrow \Natural$
satisfying $q_{\calP} \geq q_{\ent}$
and a set of $k$ quantum provers $P_{1}, \ldots, P_{k}$
of $m$-message $(q_{\calM}, q_{\calP})$-restricted $q_{\ent}$-prior-entangled
such that $(P_{1}, \ldots, P_{k}, V)$ accepts $x$ with probability at least $a(n)$,
\item[(ii)]
if $x \not\in L$, for all functions
$q'_{\calP} \colon \Nonnegative \rightarrow \Natural$
satisfying $q'_{\calP} \geq q_{\ent}$
and all sets of $k$ quantum provers $P'_{1}, \ldots, P'_{k}$
of $m$-message $(q_{\calM}, q'_{\calP})$-restricted $q_{\ent}$-prior-entangled,
$(P'_{1}, \ldots, P'_{k}, V)$ accepts $x$ with probability at most $b(n)$.
\end{itemize}
\label{Definition: QMIP(k,m,ent,a,b)}
\end{definition}

\noindent
Let $\QMIP(poly, poly, q_{\ent}, a, b)$
denote the union of the classes $\QMIP(k, m, q_{\ent}, a, b)$
over all polynomially bounded functions $k$ and $m$.
The class $\QMIP$ of languages
having quantum multi-prover interactive proof systems
is defined as follows.

\begin{definition}
A language $L$ is in $\QMIP$
iff there exists a function
$q_{\ent} \colon \Nonnegative \rightarrow \Natural$
such that, for any function
$q'_{\ent} \colon \Nonnegative \rightarrow \Natural$
satisfying $q'_{\ent} \geq q_{\ent}$,
$L$ is in $\QMIP(poly, poly, q'_{\ent}, 1, 1/2)$.
\end{definition}

Next we define the class $\QMIP^{\mathrm{(l.e.)}}$ of languages
having quantum multi-prover interactive proof systems
with at most polynomially many prior-entangled qubits.

\begin{definition}
A language $L$ is in $\QMIP^{\mathrm{(l.e.)}}$
iff there exists a polynomially bounded function
$q_{\ent} \colon \Nonnegative \rightarrow \Natural$
such that, for any polynomially bounded function
$q'_{\ent} \colon \Nonnegative \rightarrow \Natural$
satisfying $q'_{\ent} \geq q_{\ent}$,
$L$ is in $\QMIP(poly, poly, q'_{\ent}, 1, 1/2)$.
\end{definition}

Finally we define the class $\QMIP^{\mathrm{(n.e.)}}$ of languages
having quantum multi-prover interactive proof systems
without any prior entanglement.

\begin{definition}
A language $L$ is in $\QMIP^{\mathrm{(n.e.)}}$
iff $L$ is in $\QMIP(poly, poly, 0, 1, 1/2)$.
\end{definition}

\subsection{Quantum Oracle Circuits}
\label{Subsection: QOC definition}

Consider a situation in which a verifier can communicate with only one prover,
but the prover does not have his private qubits.
We call this model a {\em quantum oracle circuit\/},
since it can be regarded as a quantum counterpart
of a probabilistic oracle machine~\cite{ForRomSip94TCS, For89PhD, BabForLun91CC}
in the sense that there is no private space for the prover
during the protocol.

For the definition of quantum oracle circuits,
we use slightly different terminologies from those in the previous subsection so that they are fitted to the term `oracle' rather than `prover'.

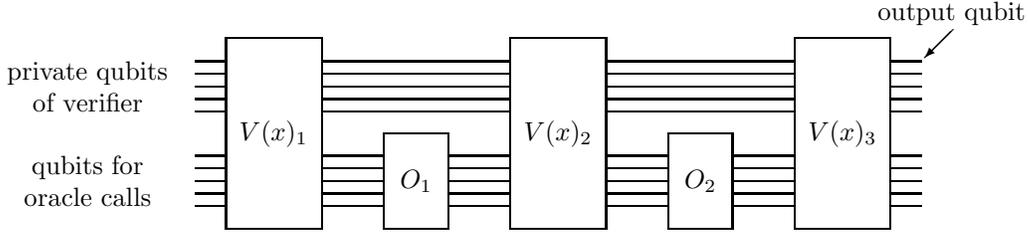
\begin{figure}[t]
\begin{center}
\setlength{\unitlength}{0.84mm}
{\small
\begin{picture}(165.75, 40)(-2, 0)
\put( -2,20){\makebox(30, 8){\shortstack{private qubits\\ of verifier}}}
\put( -2, 5){\makebox(30, 8){\shortstack{qubits for\\ oracle calls}}}
\multiput( 30,20)(0,2){5}{\line(1,0){5}}
\multiput( 30, 5)(0,2){5}{\line(1,0){5}}
\put( 35, 1.5){\framebox(15,30){$V(x)_{1}$}}
\multiput( 50,20)(0,2){5}{\line(1,0){30}}
\multiput( 50, 5)(0,2){5}{\line(1,0){10}}
\put( 60, 1.5){\framebox(10,15){$O_{1}$}}
\multiput( 70, 5)(0,2){5}{\line(1,0){10}}
\put( 80, 1.5){\framebox(15,30){$V(x)_{2}$}}
\multiput( 95,20)(0,2){5}{\line(1,0){30}}
\multiput( 95, 5)(0,2){5}{\line(1,0){10}}
\put(105, 1.5){\framebox(10,15){$O_{2}$}}
\multiput(115, 5)(0,2){5}{\line(1,0){10}}
\put(125, 1.5){\framebox(15,30){$V(x)_{3}$}}
\multiput(140,20)(0,2){5}{\line(1,0){5}}
\multiput(140, 5)(0,2){5}{\line(1,0){5}}
\put(150,33){\vector(-1,-1){4.5}}
\put(135.75,33){\makebox(28,5){output qubit}}
\end{picture}
}
\caption{Quantum circuit for a two-oracle-call quantum oracle circuit}
\label{Figure: QOCfigure}
\end{center}
\end{figure}

Given polynomially bounded functions
$m, q_{\calV}, q_{\calO} \colon \Nonnegative \rightarrow \Natural$,
an {\em $m$-oracle-call $(q_{\calV}, q_{\calO})$-restricted quantum verifier\/} $V$
for a quantum oracle circuit
is a $2m$-message $(q_{\calV}, q_{\calO})$-restricted quantum verifier
for a quantum single-prover interactive proof system.
A {\em $q_{\calO}$-restricted quantum oracle\/} $O$
for an $m$-oracle-call $(q_{\calV}, q_{\calO})$-restricted quantum verifier
is a $2m$-message $(q_{\calO}, 0)$-restricted quantum prover.
Figure~\ref{Figure: QOCfigure}
illustrates the situation of a two-oracle-call quantum oracle circuit.
Note that our definition of a quantum oracle completely
differs from the one
by Bennett, Bernstein, Brassard, and Vazirani~\cite{BenBerBraVaz97SIComp}
in which a quantum oracle is restricted to a unitary transformation
that maps $\ket{y, z}$ to $\ket{y, z \oplus f(y)}$ in one step
for an arbitrary function $f\colon \{0,1\}^{\ast} \rightarrow \{0,1\}$.

\begin{definition}
Given a polynomially bounded function
$m \colon \Nonnegative \rightarrow \Natural$
and functions $a, b \colon \Nonnegative \rightarrow [0,1]$,
a language $L$ is in $\QOC(m, a, b)$
iff there exist polynomially bounded functions
$q_{\calV}, q_{\calO} \colon \Nonnegative \rightarrow \Natural$
and an $m$-oracle-call $(q_{\calV}, q_{\calO})$-restricted quantum verifier $V$
for a quantum oracle circuit
such that, for every input $x$ of length $n$,
\begin{itemize}
\item[(i)]
if $x \in L$,
there exists
a $q_{\calO}$-restricted quantum oracle $O$ for $V$
such that $V$ with access to $O$ accepts $x$
with probability at least $a(n)$,
\item[(ii)]
if $x \not\in L$,
for all $q_{\calO}$-restricted quantum oracles $O'$ for $V$, 
$V$ with access to $O'$ accepts $x$ with probability at most $b(n)$.
\end{itemize}
\end{definition}

\noindent
Let $\QOC(poly, a, b)$ denote
the union of the classes $\QOC(m, a, b)$ over all polynomially bounded functions $m$.
The class $\QOC$ of languages
accepted by quantum oracle circuits is defined as follows.

\begin{definition}
A language $L$ is in $\QOC$
iff $L$ is in $\QOC(poly, 1, 1/2)$.
\end{definition}


\section{$\boldsymbol{\QMIP^{\mathrm{(l.e.)}} \subseteq \NEXP}$}
\label{Section: NEXP upper bound}

Now we show that
every language having a quantum multi-prover interactive proof system
is necessarily in $\NEXP$
under the assumption that
provers are allowed to share
at most polynomially many prior-entangled qubits.

A key idea of our proof is
to bound the number of private qubits of provers
without diminishing the computational power of them.
First,
in Subsection~\ref{Subsection: Single-Prover Case},
we explain our bounding technique with the single-prover case,
which is much easier to understand.
Although our result for the single-prover case
only gives the $\NEXP$ upper bound
for the class $\QIP$ of quantum single-prover interactive proofs,
it will be much of help
to understand our key idea of the proof for the multi-prover case
in Subsection~\ref{Subsection: QMIP(l.e.) subset NEXP}.
For simplicity, in this section and after,
we often drop the argument $x$ and $n$
in the various functions defined in the previous section.
We also assume that operators acting on subsystems
of a given system are extended to the entire system
by tensoring with the identity,
when it is clear from context
upon what part of a system a given operator acts.

\subsection{Single-Prover Case}
\label{Subsection: Single-Prover Case}

First we show that,
for any protocol of quantum single-prover interactive proof systems,
there exists a quantum single-prover interactive protocol
exchanging the same number of messages,
in which the prover uses only polynomially many qubits
for his private space with respect to input length,
and the probability of acceptance is exactly equal to that of the original one.
To show this, the following two theorems play very important roles.
A point of our proof is how to combine and apply these two to
the theory of quantum interactive proof systems.

\begin{theorem}[\cite{Uhl86RMathPhy, HugJozWoo93PLA}]
Let $\ket{\phi}, \ket{\psi} \in \calH_{1} \otimes \calH_{2}$ satisfy
$\tr_{\calH_{2}} \ket{\phi}\bra{\phi} = \tr_{\calH_{2}} \ket{\psi}\bra{\psi}$.
Then there is a unitary transformation $U$ over $\calH_{2}$
such that $(I_{\calH_{1}} \otimes U) \ket{\phi} = \ket{\psi}$,
where $I_{\calH_{1}}$ is the identity operator over $\calH_{1}$.
\label{Theorem: locality theorem}
\end{theorem}

\begin{theorem}[\cite{NieChu00Book}, page 110]
Let $\rho$ be a density matrix over $\calH_{1}$.
Then there exist a Hilbert space $\calH_{2}$ of $\dim(\calH_{2}) = \dim(\calH_{1})$
and a pure state $\ket{\psi} \in \calH_{1} \otimes \calH_{2}$
such that $\tr_{\calH_{2}} \ketbra{\psi} = \rho$.
\label{Theorem: purification}
\end{theorem}

Now we give a proof of our claim.

\begin{lemma}
Let $m, q_{\calV}, q_{\calM} \colon \Nonnegative \rightarrow \Natural$
be polynomially bounded functions
and $V$ be an $m$-message $(q_{\calV}, q_{\calM})$-restricted quantum verifier
for a quantum single-prover interactive proof system.
Then,
for any function $q_{\calP} \colon \Nonnegative \rightarrow \Natural$
and any $m$-message $(q_{\calM}, q_{\calP})$-restricted quantum prover $P$,
there exists an $m$-message $(q_{\calM}, q_{\calV} + q_{\calM})$-restricted
quantum prover $P'$
such that, for every input $x$,
the probability of accepting $x$ by $(P', V)$
is exactly equal to the one by $(P, V)$.
\label{Lemma: bounded-simulation for QIP}
\end{lemma}

\begin{proof}
It is assumed that
$q_{\calP} \geq q_{\calV} + q_{\calM}$,
since there is nothing to show in the case
$q_{\calP} < q_{\calV} + q_{\calM}$.
It is also assumed that the values of $m$ are even
(odd cases can be dealt with a similar argument).

Given a protocol $(P, V)$ of
an $m$-message $(q_{\calV}, q_{\calM}, q_{\calP})$-restricted
quantum single-prover interactive proof system,
we construct an $m$-message ${(q_{\calM}, q_{\calV} + q_{\calM})}$-restricted
quantum prover $P'$
such that the probability of acceptance by $(P', V)$
is exactly equal to the one by $(P, V)$ on every input.
We construct $P'$
by showing, for every input $x$,
how to construct each $P'_{j}(x)$ based on the original $P_{j}(x)$.
In the following proof,
each $P_{j}(x)$ and $P'_{j}(x)$ will be abbreviated as $P_{j}$ and $P'_{j}$,
respectively.

Let $\calP' = l_{2}(\Sigma^{q_{\calM} + q_{\calV}})$
be the Hilbert space corresponding to the private qubits of $P'$.
Let each
$\ket{\psi_{j}}, \ket{\phi_{j}} \in \calV \otimes \calM \otimes \calP$,
for $1 \leq j \leq m/2$,
denote a state of the original
$m$-message $(q_{\calV}, q_{\calM}, q_{\calP})$-restricted
quantum interactive proof system
defined in a recursive manner by
\[
\begin{array}{lcll}
\ket{\phi_{1}} & = & V_{1} \ket{\init}, & \\
\ket{\phi_{j}}
& = & V_{j} P_{j-1} \ket{\phi_{j-1}}, & 2 \leq j \leq m/2,\\
\ket{\psi_{j}} & = & P_{j} \ket{\phi_{j}}, & 1 \leq j \leq m/2.
\end{array}
\]
Here $\ket{\init} \in {\calV} \otimes {\calM} \otimes {\calP}$
is the initial state
in which all the qubits are the $\ket{0}$-states.
Notice that
$\tr_{\calM \otimes \calP} \ketbra{\psi_{j}}
= \tr_{\calM \otimes \calP} \ketbra{\phi_{j}}$
for each $1 \leq j \leq m/2$,
since each $P_{j}$ acts only on the qubits in $\calM \otimes \calP$.

From Theorem~\ref{Theorem: purification},
there exist states
$\ket{\psi'_{j}}, \ket{\phi'_{j}} \in \calV \otimes \calM \otimes \calP'$
such that
\[
\begin{array}{lcl}
\tr_{\calP'} \ketbra{\phi'_{j}} & = & \tr_{\calP} \ketbra{\phi_{j}},\\
\tr_{\calP'} \ketbra{\psi'_{j}} & = & \tr_{\calP} \ketbra{\psi_{j}},
\end{array}
\]
for each $1 \leq j \leq m/2$.
Thus we have
\[
\tr_{\calM \otimes \calP'} \ketbra{\psi'_{j}}
= \tr_{\calM \otimes \calP} \ketbra{\psi_{j}}
= \tr_{\calM \otimes \calP} \ketbra{\phi_{j}}
= \tr_{\calM \otimes \calP'} \ketbra{\phi'_{j}},
\]
for each $1 \leq j \leq m/2$.

Therefore, by Theorem~\ref{Theorem: locality theorem},
there exists a unitary transformation $P'_{j}$
acting on $\calM \otimes \calP'$
such that $P'_{j} \ket{\phi'_{j}} = \ket{\psi'_{j}}$
for each $1 \leq j \leq m/2$.

Having defined $P'_{j}$, $\ket{\phi'_{j}}$, and $\ket{\psi'_{j}}$
for each $1 \leq j \leq m/2$,
compare the state just before the final measurement is performed
in the original protocol and that in the constructed protocol.
Let $\ket{\phi_{m/2+1}} = V_{m/2+1} \ket{\psi_{m/2}}$
and $\ket{\phi'_{m/2+1}} = V_{m/2+1} \ket{\psi'_{m/2}}$.
These $\ket{\phi_{m/2+1}}$ and $\ket{\phi'_{m/2+1}}$
are exactly the states we want to compare.
Noticing that
$\tr_{\calP} \ketbra{\psi_{m/2}}
= \tr_{\calP'} \ketbra{\psi'_{m/2}}$,
we have
$\tr_{\calP} \ketbra{\phi_{m/2+1}} = \tr_{\calP} \ketbra{\phi'_{m/2+1}}$,
since $V_{m/2+1}$ acts only on $\calV \otimes \calM$.
This implies that
the verifier $V$ cannot distinguish $\ket{\phi'_{m/2+1}}$ from $\ket{\phi_{m/2+1}}$ at all.
Hence, for every input $x$,
the probability of accepting $x$ in the protocol $(P', V)$
is exactly equal to the one in the original protocol $(P, V)$.
Thus we have the assertion.
\end{proof}

\subsection{$\boldsymbol{\QMIP^{\mathrm{(l.e.)}} \subseteq \NEXP}$}
\label{Subsection: QMIP(l.e.) subset NEXP}

In the proof of Lemma~\ref{Lemma: bounded-simulation for QIP}
we decomposed the Hilbert space of the proof system
into $\calV \otimes \calM$ and $\calP$
and used Theorem~\ref{Theorem: purification}
by taking $\calV \otimes \calM$
as a Hilbert space $\calH_{1}$ of Theorem~\ref{Theorem: purification}.
For $k$-prover cases, however,
if we focus on one fixed prover $P_{i}$
and decompose the Hilbert space of the proof system
into the private space of $P_{i}$ and the rest,
Theorem~\ref{Theorem: purification} is of no help,
because the number of qubits of the proof system
out of $\calP_{i}$ may be no longer bounded polynomial in input length.
Instead of Theorem~\ref{Theorem: purification},
we show the following theorem, which is useful even for $k$-prover cases.

\begin{theorem}
Fix a state $\ket{\phi}$
in $\calH_{1} \otimes \calH_{2} \otimes \calH_{3}$
and a unitary transformation $U$
over $\calH_{2} \otimes \calH_{3}$ arbitrarily,
and let $\ket{\psi}$ denote
$(I_{\calH_{1}} \otimes U) \ket{\phi}$.
Then, for any Hilbert space $\calH'_{3}$ of
$\dim(\calH'_{3}) \leq \dim(\calH_{3})$
such that
there is a state
$\ket{\phi'}$ in $\calH_{1} \otimes \calH_{2} \otimes \calH'_{3}$
satisfying
$\tr_{\calH'_{3}} \ketbra{\phi'} = \tr_{\calH_{3}} \ketbra{\phi}$,
there exist
a Hilbert space $\calH''_{3}$ of
$\dim(\calH''_{3}) = (\dim(\calH_{2}))^{2} \cdot \dim(\calH'_{3})$
and a state $\ket{\psi'}$ in
$\calH_{1} \otimes \calH_{2} \otimes \calH''_{3}$
such that
$\tr_{\calH''_{3}} \ketbra{\psi'} = \tr_{\calH_{3}} \ketbra{\psi}$.
\label{Theorem: bounding theorem}
\end{theorem}

For the proof of Theorem \ref{Theorem: bounding theorem}, 
we use the entanglement measure introduced by Nielsen~\cite{Nie01QIP}.
Let us decompose
a vector $\ket{\xi} \in \calH_{1} \otimes \calH_{2}$ into
\begin{equation}
\ket{\xi}=\sum_{i,j}\alpha_{ij}\ket{e_{i}^{1}}\otimes\ket{e_{j}^{2}},
\label{Equation: kg1}
\end{equation}
where $\{ \ket{e_{i}^{1}} \}$ and $\{ \ket{e_{i}^{2}} \}$
are orthonormal bases
of $\calH_{1}$ and $\calH_{2}$, respectively.
Then the {\em entanglement measure\/}
$\ent_{2}(\ket{\xi}, \calH_1, \calH_2)$
is defined by
the minimum number of non-zero terms in the right hand side 
of (\ref{Equation: kg1}),
where the minimum is taken over all the possible choices of 
the bases $\{ \ket{e_{i}^{1}} \}$ and $\{ \ket{e_{i}^{2}} \}$.
The decomposition with the minimum number of non-zero terms
is given by the Schmidt decomposition~\cite{Uhl86RMathPhy},
\[
\ket{\xi} = \sum_{i} \beta_{i} \ket{e_{i}^{1}} \otimes \ket{e_{i}^{2}},
\]
where each $\ket{e_{i}^{1}}$ and $\ket{e_{i}^{2}}$
is a normalized eigenvector of
$\tr_{\calH_{1}} \ketbra{\xi}$ and  
$\tr_{\calH_{2}} \ketbra{\xi}$, respectively. 
Therefore, 
the entanglement measure $\ent_2(\ket{\xi}, \calH_1, \calH_2)$
is nothing but
the minimum dimension of the Hilbert space $\calH'_2$
such that there is a vector
$\ket{\xi'} \in \calH_1 \otimes \calH'_2$
that satisfies
$\tr_{\calH_2} \ketbra{\xi} = \tr_{\calH'_2} \ketbra{\xi'}$.

We extend the definition of $\ent_2(\cdot, \cdot, \cdot)$ to a three-party case.
For a vector
$\ket{\zeta} \in \calH_{1} \otimes \calH_{2} \otimes \calH_{3}$,
define the {\em three-party entanglement measure\/} 
$\ent_{3}(\ket{\zeta}, \calH_1, \calH_2, \calH_3)$
as the minimum number of non-zero terms in the decomposition
\[
\ket{\zeta} = \sum_{i,j,k} \gamma_{ijk}
                           \ket{e^1_i} \otimes \ket{e^2_j} \otimes \ket{e^3_k},
\]
where $\{ \ket{e^j_i} \}$ denotes an orthonormal basis
of the space $\calH_j$ for each $j = 1, 2, 3$.

\begin{proofof}{Theorem \ref{Theorem: bounding theorem}}
Since 
$\ent_{2}(\ket{\psi}, \calH_1 \otimes \calH_2, \calH_3)$
gives
the minimum dimension of $\calH''_{3}$
such that there is a state
$\ket{\psi'} \in \calH_1 \otimes \calH_2 \otimes \calH''_{3}$
satisfying
$\tr_{\calH''_3} \ketbra{\psi'} = \tr_{\calH_3} \ketbra{\psi}$,
it is sufficient to show that
$\ent_{2}(\ket{\psi}, \calH_1 \otimes \calH_2, \calH_3)
\leq \dim(\calH'_3) \cdot (\dim(\calH_2))^2$.
This can be proved as follows:
\begin{eqnarray*}
\ent_{2}(\ket{\psi}, \calH_1 \otimes \calH_2, \calH_3)
& \leq &
\ent_{3}(\ket{\psi}, \calH_1, \calH_2, \calH_3)\\
& \leq &
\ent_{3}(\ket{\phi}, \calH_1, \calH_2, \calH_3)
      \cdot \dim(\calH_2)\\
& \leq &
\ent_{2}(\ket{\phi}, \calH_1 \otimes \calH_2, \calH_3)
      \cdot (\dim(\calH_2))^2\\
& \leq &
\dim(\calH'_3)
      \cdot (\dim(\calH_2))^2.
\end{eqnarray*}
The first inequality directly comes from the definition of
the entanglement measure.
To prove the second and third inequalities,
let
$\ket{\phi}
  = \sum_{i,j,k} \gamma_{ijk} 
                   \ket{e^1_i} \otimes \ket{e^2_j} \otimes \ket{e^3_k}$
be the decomposition of $\ket{\phi}$
with respect to the orthonormal bases
$\{ \ket{e^1_i} \}$, $\{ \ket{e^2_i} \}$, and $\{ \ket{e^3_i} \}$
of $\calH_{1}$, $\calH_{2}$, and $\calH_{3}$, respectively,
and let
$\ket{\phi} = \sum_{i} \beta_{i} \ket{f^{1,2}_i} \otimes \ket{f^3_i}$
be that of $\ket{\phi}$
with respect to the orthonormal bases
$\{ \ket{f^{1,2}_i} \}$ and $\{ \ket{f^3_i} \}$
of $\calH_{1} \otimes \calH_{2}$ and $\calH_{3}$, respectively.
The second and third inequalities are the consequences of 
the equality 
\[
\ket{\psi}
=
 \sum_{i,j,k} \gamma_{ijk} 
                \ket{e^1_i} \otimes U(\ket{e^2_j}\otimes\ket{e^3_k})
=
 \sum_{i,j,k} \gamma_{ijk} 
                \ket{e^1_i} \otimes
                \left(
                  \sum_{l=1}^{\dim(\calH_2)} \beta'_{jkl} 
                  \ket{e^{2}_{jkl}} \otimes \ket{e^{3}_{jkl}}
                \right)
\]
and the equality
\[
\ket{\phi}
=
\sum_{i} \beta_{i} \ket{f^{1,2}_i} \otimes \ket{f^3_i}
=
\sum_{i} \beta_{i} \left(
                     \sum_{j=1}^{\dim(\calH_2)}
                       \beta''_{ij} \ket{f^1_{ij}} \otimes \ket{f^2_{ij}}
                   \right)
                   \otimes \ket{f^3_j},
\]
respectively,
where 
$\sum_{l=1}^{\dim(\calH_2)} \beta'_{jkl} \ket{e^{2}_{jkl}}
\otimes \ket{e^{3}_{jkl}}$
and
$\sum_{j=1}^{\dim(\calH_2)} \beta''_{ij} \ket{f^1_{ij}} \otimes \ket{f^2_{ij}}$
are the Schmidt decompositions of $U(\ket{e^2_j} \otimes \ket{e^3_k})$
and $\ket{f^{1,2}_i}$, respectively.
The fourth inequality is from
the definition of the entanglement measure,
which ensures that
$\ent_2(\ket{\phi}, \calH_1 \otimes \calH_2, \calH_3)
\leq \dim(\calH'_3)$
holds.
\end{proofof}

Now we are ready to show the following lemma.

\begin{lemma}
Let $k, m, q_{\calV}, q_{\calM}, q_{\ent} \colon \Nonnegative \rightarrow \Natural$
be polynomially bounded functions
and $V$ be an $m$-message $(q_{\calV}, q_{\calM})$-restricted quantum verifier
for a quantum $k$-prover interactive proof system.
Then,
for any function $q_{\calP} \colon \Nonnegative \rightarrow \Natural$
satisfying $q_{\calP} \geq q_{\ent}$
and any set of $m$-message $(q_{\calM}, q_{\calP})$-restricted
$q_{\ent}$-prior-entangled quantum provers $P_{1}, \ldots, P_{k}$,
there exists a set of $m$-message
$(q_{\calM}, q_{\ent} + {2 \lfloor m/2 + 1/2 \rfloor q_{\calM}})$-restricted
$q_{\ent}$-prior-entangled quantum provers $P'_{1}, \ldots, P'_{k}$
such that, for every input $x$,
the probability of accepting $x$ by $(P'_{1}, \ldots, P'_{k}, V)$
is exactly equal to the one by $(P_{1}, \ldots, P_{k}, V)$.
\label{Lemma: bounded-simulation for QMIP}
\end{lemma}

\begin{proof}
It is assumed that
$q_{\calP} \geq q_{\ent} + {2 \lfloor m/2 + 1/2 \rfloor q_{\calM}}$,
since there is nothing to show in the case
$q_{\calP} < q_{\ent} + {2 \lfloor m/2 + 1/2 \rfloor q_{\calM}}$.
It is also assumed that the values of $m$ are even,
and thus ${2 \lfloor m/2 + 1/2 \rfloor q_{\calM}} = m q_{\calM}$
(odd cases can be dealt with a similar argument).

Given a protocol $(P_{1}, \ldots, P_{k}, V)$ of
an $m$-message $(q_{\calV}, q_{\calM}, q_{\calP})$-restricted
$q_{\ent}$-prior-entangled quantum $k$-prover interactive proof system,
we first show that $P_{1}$ can be replaced by
an $m$-message
$(q_{\calM}, q_{\ent} + m q_{\calM})$-restricted
$q_{\ent}$-prior-entangled quantum prover $P'_{1}$
such that the probability of acceptance by
$(P'_{1}, P_{2}, \ldots, P_{k}, V)$
is exactly equal to the one by $(P_{1}, \ldots, P_{k}, V)$ on every input.
Having shown this, we repeat the same process for each of provers
to construct a protocol $(P'_{1}, P'_{2}, P_{3}, \ldots, P_{k}, V)$
from $(P'_{1}, P_{2}, P_{3}, \ldots, P_{k}, V)$ and so on,
and finally we obtain a protocol $(P'_{1}, \ldots, P'_{k}, V)$
in which all of $P'_{1}, \ldots, P'_{k}$ are
$m$-message
$(q_{\calM}, q_{\ent} + m q_{\calM})$-restricted
$q_{\ent}$-prior-entangled quantum provers.
We construct $P'_{1}$
by showing, for every input $x$,
how to construct each $P'_{1,j}(x)$ based on the original $P_{1,j}(x)$.
In the following proof,
each $P_{i,j}(x)$ and $P'_{i,j}(x)$ will be abbreviated as
$P_{i,j}$ and $P'_{i,j}$, respectively.

Let each
$\ket{\psi_{j}}, \ket{\phi_{j}}
\in \calV
\otimes \calM_{1} \otimes \cdots \otimes \calM_{k}
\otimes \calP_{1} \otimes \cdots \otimes \calP_{k}$,
for $1 \leq j \leq m/2$,
denote a state of the original
$m$-message $(q_{\calV}, q_{\calM}, q_{\calP})$-restricted
$q_{\ent}$-prior-entangled quantum $k$-prover interactive proof system
defined in a recursive manner by
\[
\begin{array}{lcll}
\ket{\phi_{1}} & = & V_{1} \ket{\init}, & \\
\ket{\phi_{j}}
& = & V_{j} P_{k,j-1} \cdots P_{1,j-1} \ket{\phi_{j-1}}, & 2 \leq j \leq m/2,\\
\ket{\psi_{j}} & = & P_{1,j} \ket{\phi_{j}}, & 1 \leq j \leq m/2.
\end{array}
\]
Here $\ket{\init}
\in \calV
\otimes \calM_{1} \otimes \cdots \otimes \calM_{k}
\otimes \calP_{1} \otimes \cdots \otimes \calP_{k}$
is the initial state
in which the first $q_{\ent}(n)$ qubits in each $\calP_{j}$
may be entangled with private qubits of other provers than $P_{j}$.
All the qubits other than these prior-entangled qubits
are the $\ket{0}$-states in the state $\ket{\init}$.
Note that
$\tr_{\calM_{1} \otimes \calP_{1}} \ketbra{\psi_{j}}
= \tr_{\calM_{1} \otimes \calP_{1}} \ketbra{\phi_{j}}$,
for each $1 \leq j \leq m/2$.

We define each $P'_{1,j}$ recursively.
To define $P'_{1,1}$, consider the states $\ket{\phi_{1}}$ and $\ket{\psi_{1}}$.
Let $\ket{\phi'_{1}} = \ket{\phi_{1}}$.
Since all of the last $(q_{\calP} - q_{\ent})$ qubits
in $\calP_{1}$ in the state $\ket{\phi_{1}}$
are the $\ket{0}$-states
and $\ket{\psi_{1}} = P_{1,1}\ket{\phi_{1}}$,
by Theorem~\ref{Theorem: bounding theorem},
there exists a state $\ket{\psi'_{1}}$
in $\calV
\otimes \calM_{1} \otimes \cdots \otimes \calM_{k}
\otimes \calP_{1} \otimes \cdots \otimes \calP_{k}$
such that
\[
\tr_{\calP_{1}} \ketbra{\psi'_{1}} = \tr_{\calP_{1}} \ketbra{\psi_{1}}
\]
and all but the first $q_{\ent} + 2q_{\calM}$ qubits
in $\calP_{1}$ are the $\ket{0}$-states in the state $\ket{\psi'_{1}}$. 
Furthermore we have
\[
\tr_{\calM_{1} \otimes \calP_{1}} \ketbra{\psi'_{1}}
= \tr_{\calM_{1} \otimes \calP_{1}} \ketbra{\psi_{1}}
= \tr_{\calM_{1} \otimes \calP_{1}} \ketbra{\phi_{1}}
= \tr_{\calM_{1} \otimes \calP_{1}} \ketbra{\phi'_{1}}.
\]
Therefore, by Theorem~\ref{Theorem: locality theorem},
there exists a unitary transformation $Q_{1,1}$
acting on $\calM_{1} \otimes \calP_{1}$
such that $Q_{1,1} \ket{\phi'_{1}} = \ket{\psi'_{1}}$
and $Q_{1,1}$ is of the form
$P'_{1,1} \otimes I_{q_{\calP} - q_{\ent} - m q_{\calM}}$,
where $P'_{1,1}$ is a unitary transformation acting on
qubits in $\calM_{1}$ and the first $q_{\ent} + m q_{\calM}$ qubits of $\calP_{1}$,
and $I_{q_{\calP} - q_{\ent} - m q_{\calM}}$ is
the $(q_{\calP} - q_{\ent} - m q_{\calM})$-dimensional identity matrix.

Assume that $Q_{1,j}$, $\ket{\phi'_{j}}$, and $\ket{\psi'_{j}}$
have been defined for each $j$, $1 \leq j \leq \xi \leq m/2-1$,
to satisfy
\begin{itemize}
\item
\begin{tabbing}
$\ket{\phi'_{1}}$ \= $=$ \= $V_{1} \ket{\init}$,\\
$\ket{\phi'_{j}}$
\> $=$ \> $V_{j} P_{k,j-1} \cdots P_{2,j-1}Q_{1,j-1} \ket{\phi'_{j-1}}$,
\= $2 \leq j \leq \xi$,\\
$\ket{\psi'_{j}}$ \> $=$ \> $Q_{1,j} \ket{\phi'_{j}},$ \> $1 \leq j \leq \xi$.
\end{tabbing}
\item
$\tr_{\calP_{1}} \ketbra{\psi_{j}}
= \tr_{\calP_{1}} \ketbra{\psi'_{j}}$, $1 \leq j \leq \xi$.
\item
All but the first $q_{\ent} + 2(j-1) q_{\calM}$ qubits
in $\calP_{1}$ are the $\ket{0}$-states in the state $\ket{\phi'_{j}}$.
\item
All but the first $q_{\ent} + 2j q_{\calM}$ qubits
in $\calP_{1}$ are the $\ket{0}$-states in the state $\ket{\psi'_{j}}$.
\end{itemize}
Notice that $Q_{1,1}$, $\ket{\phi'_{1}}$, and $\ket{\psi'_{1}}$ defined above
satisfy such conditions.
Define $Q_{1,\xi+1}$, $\ket{\phi'_{\xi+1}}$, and $\ket{\psi'_{\xi+1}}$
in the following way
to satisfy the above four conditions for $j = \xi+1$.

Let $U_{\xi} = V_{\xi+1} P_{k,\xi} \cdots P_{2,\xi}$
and define
$\ket{\phi'_{\xi+1}} = U_{\xi} \ket{\psi'_{\xi}}$.
Then
all but the first $q_{\ent} + 2\xi q_{\calM}$ qubits
in $\calP_{1}$ are the $\ket{0}$-states in the state $\ket{\phi'_{\xi+1}}$,
since none of $P_{2,\xi}, \ldots, P_{k,\xi}, V_{\xi+1}$
acts on the space $\calP_{1}$
and $\ket{\psi'_{\xi}}$ satisfies the fourth condition.
Since
$\tr_{\calP_{1}} \ketbra{\psi_{\xi}}
= \tr_{\calP_{1}} \ketbra{\psi'_{\xi}}$,
by Theorem~\ref{Theorem: locality theorem},
there exists a unitary transformation $A_{\xi}$
acting on $\calP_{1}$
such that $A_{\xi} \ket{\psi'_{\xi}} = \ket{\psi_{\xi}}$.
Thus we have
\begin{equation}
\ket{\psi_{\xi+1}}
=
P_{1,\xi+1} U_{\xi} \ket{\psi_{\xi}}
=
P_{1,\xi+1} U_{\xi} A_{\xi} \ket{\psi'_{\xi}}
=
P_{1,\xi+1} A_{\xi} U_{\xi} \ket{\psi'_{\xi}}
=
P_{1,\xi+1} A_{\xi} \ket{\phi'_{\xi+1}}.
\label{Equation: psiphi'}
\end{equation}
Hence, by Theorem~\ref{Theorem: bounding theorem},
there exists a state $\ket{\psi'_{\xi+1}}$
such that
\begin{equation}
\tr_{\calP_{1}} \ketbra{\psi'_{\xi+1}}
= \tr_{\calP_{1}} \ketbra{\psi_{\xi+1}}
\label{Equation: trP1psi}
\end{equation}
and all but the first $q_{\ent} + 2(\xi+1) q_{\calM}$ qubits
in $\calP_{1}$ are the $\ket{0}$-states in the state $\ket{\psi'_{\xi+1}}$.
From (\ref{Equation: psiphi'}) and (\ref{Equation: trP1psi}), we have
\[
\tr_{\calM_{1} \otimes \calP_{1}} \ketbra{\psi'_{\xi+1}}
=
\tr_{\calM_{1} \otimes \calP_{1}} \ketbra{\psi_{\xi+1}}
=
\tr_{\calM_{1} \otimes \calP_{1}} \ketbra{\phi'_{\xi+1}},
\]
since $P_{1,\xi+1}$ and $A_{\xi}$ act only on $\calM_{1} \otimes \calP_{1}$.
Therefore, by Theorem~\ref{Theorem: locality theorem},
there exists a unitary transformation $Q_{1,\xi+1}$
acting on $\calM_{1} \otimes \calP_{1}$
such that $Q_{1,\xi+1} \ket{\phi'_{\xi+1}} = \ket{\psi'_{\xi+1}}$.
It follows that
$Q_{1,\xi+1}$ is of the form
$P'_{1,\xi+1} \otimes I_{q_{\calP} - q_{\ent} - m q_{\calM}}$,
where $P'_{1,\xi+1}$ is a unitary transformation acting on
qubits in $\calM_{1}$ and the first $q_{\ent} + m q_{\calM}$ qubits of $\calP_{1}$,
because all of the last $q_{\calP} - q_{\ent} - m q_{\calM}$ qubits
in $\calP_{1}$ are the $\ket{0}$-states
in both of the states $\ket{\phi'_{\xi+1}}$ and $\ket{\psi'_{\xi+1}}$.
One can see that $Q_{1,\xi+1}, \ket{\phi'_{\xi+1}}$, and $\ket{\psi'_{\xi+1}}$
satisfy the four conditions above by their construction.

Having defined $Q_{1,j}, \ket{\phi'_{j}}, \ket{\psi'_{j}}$
for each $1 \leq j \leq m/2$,
compare the state just before the final measurement is performed
in the original protocol and that in the modified protocol
applying $Q_{1,j}$'s instead of $P_{1,j}$'s.
For $U_{m/2} = V_{m/2+1} P_{k,m/2} \cdots P_{2,m/2}$,
let $\ket{\phi_{m/2+1}} = U_{m/2} \ket{\psi_{m/2}}$
and $\ket{\phi'_{m/2+1}} = U_{m/2} \ket{\psi'_{m/2}}$.
These $\ket{\phi_{m/2+1}}$ and $\ket{\phi'_{m/2+1}}$
are exactly the states we want to compare.
Noticing that
$\tr_{\calP_{1}} \ketbra{\psi_{m/2}}
= \tr_{\calP_{1}} \ketbra{\psi'_{m/2}}$,
we have
$\tr_{\calP_{1}} \ketbra{\phi_{m/2+1}} = \tr_{\calP_{1}} \ketbra{\phi'_{m/2+1}}$,
since none of $P_{2,m/2}, \ldots, P_{k,m/2}, V_{m/2+1}$
acts on $\calP_{1}$.
Thus we have
\[
\tr_{\calP_{1} \otimes \cdots \otimes \calP_{k}} \ketbra{\phi_{m/2+1}}
= \tr_{\calP_{1} \otimes \cdots \otimes \calP_{k}} \ketbra{\phi'_{m/2+1}},
\]
which implies that
the verifier $V$ cannot distinguish $\ket{\phi'_{m/2+1}}$ from $\ket{\phi_{m/2+1}}$ at all.
Hence, for every input $x$,
the probability of accepting $x$ in the protocol $(Q_{1}, P_{2}, \ldots, P_{k}, V)$
is exactly equal to the one in the original protocol $(P_{1}, \ldots, P_{k}, V)$,
and $Q_{1}$ uses only $q_{\ent} + m q_{\calM}
 = q_{\ent} + 2 \cdot (m/2) \cdot q_{\calM}$ qubits in his private space.
In the protocol $(Q_{1}, P_{2}, \ldots, P_{k}, V)$,
each $Q_{1,j}$ is described as
$Q_{1,j} = P'_{1,j} \otimes I_{q_{\calP} - q_{\ent} - m q_{\calM}}$,
where $P'_{1,\xi+1}$ is a unitary transformation acting on
qubits in $\calM_{1}$ and the first $q_{\ent} + m q_{\calM}$ qubits of $\calP_{1}$.
Consequently, by constructing
an $m$-message $(q_{\calM}, q_{\ent} + m q_{\calM})$-restricted
quantum prover $P'_{1}$ from each $P'_{1,j}$,
for every input $x$,
the probability of accepting $x$ in the protocol $(P'_{1}, P_{2}, \ldots, P_{k}, V)$
is exactly equal to the one in the original protocol $(P_{1}, \ldots, P_{k}, V)$.

Now we repeat the above process for each of provers,
and finally we obtain a protocol $(P'_{1}, \ldots, P'_{k}, V)$
in which all $k$ provers
are $m$-message ${(q_{\calM}, q_{\ent} + m q_{\calM})}$-restricted
quantum provers.
It is obvious that,
for every input $x$, the probability of accepting $x$ in the protocol $(P'_{1}, \ldots, P'_{k}, V)$
is exactly equal to the one in the original protocol $(P_{1}, \ldots, P_{k}, V)$, and we have the assertion.
\end{proof}

From Lemma~\ref{Lemma: bounded-simulation for QMIP},
it is straightforward to show the following lemma.

\begin{lemma}
For any polynomially bounded functions
$k, m, q_{\ent} \colon \Nonnegative \rightarrow \Natural$,
$\QMIP(k, m, q_{\ent}, 1, 1/2) \subseteq \NEXP$.
\label{Lemma: NEXP upper bound}
\end{lemma}

\begin{proof}
For convenience,
we assume that the values of $m$ are even
(odd cases can be dealt with a similar argument).

Let $L$ be a language in $\QMIP(k, m, q_{\ent}, 1, 1/2)$.
Then, from Definition~\ref{Definition: QMIP(k,m,ent,a,b)}
together with Lemma~\ref{Lemma: bounded-simulation for QMIP},
there exist polynomially bounded functions
$q_{\calV}, q_{\calM} \colon \Nonnegative \rightarrow \Natural$
and an $m$-message $(q_{\calV}, q_{\calM})$-restricted quantum verifier $V$
for a quantum $k$-prover interactive proof system
such that, for every input $x$,
(i) if $x$ is in $L$,
there exists a set of $k$ quantum provers $P_{1}, \ldots, P_{k}$
of $m$-message $(q_{\calM}, q_{\ent} + m q_{\calM})$-restricted $q_{\ent}$-prior-entangled
such that $(P_{1}, \ldots, P_{k}, V)$ accepts $x$ with certainty,
and (ii) if $x$ is not in $L$,
for all sets of $k$ quantum provers $P'_{1}, \ldots, P'_{k}$
of $m$-message $(q_{\calM}, q_{\ent} + m q_{\calM})$-restricted $q_{\ent}$-prior-entangled,
$(P'_{1}, \ldots, P'_{k}, V)$ accepts $x$ with probability at most $1/2$.

For an input $x$ of length $n$,
consider a classical simulation
of this quantum $k$-prover interactive proof system
by a non-deterministic Turing machine.
Let $p_{1}$ be arbitrary fixed polynomial.
First, for the initial state $\ket{\init}$,
an approximation $\ket{\widetilde{\psi}_{\mathrm{init}}}$
of $\ket{\init}$ can be guessed
in time non-deterministic exponential in $n$
with accuracy of
$\norm{\ket{\widetilde{\psi}_{\mathrm{init}}} - \ket{\init}} < 2^{-p_{1}(n)}$.
Next, since each $V_{j}$ applied in the original proof system
is polynomial-time uniformly generated and
$q_{\calV}$ and $q_{\calM}$ are polynomially bounded functions,
it is routine to show that
an approximation $\widetilde{V}_{j}$ of a matrix description of $V_{j}$
can be computed in time exponential in $n$ with accuracy of
$\norm{\widetilde{V}_{j} - V_{j}} < 2^{-p_{1}(n)}$.
Finally,
since $q_{\calM}$ and $q_{\calP} = q_{\ent} + m q_{\calM}$
are polynomially bounded functions,
for each operation $P_{i,j}$ of the $i$th prover
applied in the original proof system,
an approximation $\widetilde{P}_{i,j}$ of a matrix description of $P_{i,j}$
can be guessed in time non-deterministic exponential in $n$
with accuracy of
$\norm{\widetilde{P}_{i,j} - P_{i,j}} < 2^{-p_{1}(n)}$.
Thus, for the quantum state
\[
\ket{\psi_{\mathrm{final}}} =
V_{m/2+1} P_{k,m/2} \cdots P_{1,m/2} V_{m/2} \cdots P_{k,1} \cdots P_{1,1} V_{1} \ket{\init},
\]
which is the state just before the final measurement in the proof system,
the approximation $\ket{\widetilde{\psi}_{\mathrm{final}}}$ of $\ket{\final}$
can be computed in time non-deterministic exponential in $n$
with accuracy of
${\norm{\ket{\widetilde{\psi}_{\mathrm{final}}} - \ket{\final}}} < 2^{-p_{2}(n)}$
for any fixed polynomial $p_{2}$ by appropriately choosing $p_{1}$.

Now, after having computed $\ket{\widetilde{\psi}_{\mathrm{final}}}$,
a measurement of the output qubit is simulated
by summing up squares of the computed amplitudes in the accepting states.
The input $x$ is accepted if and only if this sum,
the computed probability that the measurement results in $\ket{1}$,
is more than $1 -  \varepsilon$.
From the property of the original proof system,
this computed probability is more than $1 - 2^{-2p_{2}(n)}$ if $x$ is in $L$,
while it is less than $1/2 + 2^{-2p_{2}(n)}$ if $x$ is not in $L$.
Thus, taking $p_{2} = n$ and $\varepsilon = 2^{-2n}$,
the input $x$ is accepted if and only if $x$ is in $L$
and the whole computation is done in time non-deterministic exponential in $n$.
\end{proof}

Hence we have the following theorem.

\begin{theorem}
$\QMIP^{\mathrm{(l.e.)}} \subseteq \NEXP$.
\label{Theorem: QMIP(l.e.) subseteq NEXP}
\end{theorem}

Note that our upper bound of $\NEXP$
holds even if we allow protocols with two-sided bounded error,
since the proof of Lemma~\ref{Lemma: bounded-simulation for QMIP}
does not depend on the accepting probabilities $a, b$,
and the proof of Lemma~\ref{Lemma: NEXP upper bound}
can be easily modified to two-sided bounded error cases.


\section{$\boldsymbol{\QMIP^{\mathrm{(n.e.)}} = \QOC = \NEXP}$}
\label{Section: QMIP(n.e.) = QOC = NEXP}

In the previous section, we proved that
the class of languages
having quantum multi-prover interactive proof systems
is necessarily contained in $\NEXP$
under the assumption that
provers are allowed to share
at most polynomially many prior-entangled qubits.
As a special case of this,
it is proved in this section that,
if provers do not share any prior entanglement with each other,
the class of languages
having quantum multi-prover interactive proof systems is equal to $\NEXP$.
Another result related to this is that $\QOC$ is also equal to $\NEXP$,
or in other words,
the class of languages
having quantum single-prover interactive proof systems
is also equal to $\NEXP$
if a prover does not have his private qubits.

The inclusions $\QMIP^{\mathrm{(n.e.)}} \subseteq \NEXP$
and $\QOC \subseteq \NEXP$ directly come from
Lemma~\ref{Lemma: NEXP upper bound}.
Thus it is sufficient for our claim
to show $\NEXP \subseteq \QMIP^{\mathrm{(n.e.)}} \subseteq \QOC$.
Fortunately,
in the cases without prior entanglement,
it is easy to show that
a quantum verifier
can successfully simulate any classical multi-prover protocol,
in particular,
a one-round two-prover classical interactive protocol
that can verify a language in $\NEXP$
with exponentially small one-sided error~\cite{FeiLov92STOC}.
Thus, we have the following theorem and corollary.
The proof of Theorem~\ref{Theorem: NEXP subseteq QMIP(n.e.)}
is straightforward, and thus omitted here
(see Appendix~\ref{Appendix: NEXP subseteq QMIP(n.e.)}).

\begin{theorem}
$\NEXP \subseteq \QMIP^{\mathrm{(n.e.)}}$.
\label{Theorem: NEXP subseteq QMIP(n.e.)}
\end{theorem}

\begin{corollary}
For prior unentangled cases,
if a language $L$ has
a quantum multi-prover interactive proof system with two-sided bounded error,
then $L$ has a two-message quantum two-prover interactive proof system
with exponentially small one-sided error.
\end{corollary}

The remainder of this section is devoted to the proof of
$\QMIP^{\mathrm{(n.e.)}} \subseteq \QOC$.

\begin{lemma}
Let
$k, m \colon \Nonnegative \rightarrow \Natural$
be polynomially bounded functions,
and $a, b \colon \Nonnegative \rightarrow [0,1]$ be functions
satisfying $a \geq b$.
Then
$\QMIP(k, m, 0, a, b)
\subseteq
\QOC(k \lfloor (m+1)/2 \rfloor, a, b)$.
\label{Lemma: QOC simulation of QMIP(n.e.)}
\end{lemma}

\begin{proof}
For simplicity, we assume that the values of $m$ are even,
and thus $k \lfloor (m+1)/2 \rfloor = km/2$
(odd cases can be proved with a similar argument).

Let $L$ be a language in $\QMIP(k, m, 0, a, b)$.
Then, from Definition~\ref{Definition: QMIP(k,m,ent,a,b)}
together with Lemma~\ref{Lemma: bounded-simulation for QMIP},
there exist polynomially bounded functions
$q_{\calV}, q_{\calM} \colon \Nonnegative \rightarrow \Natural$
and an $m$-message $(q_{\calV}, q_{\calM})$-restricted quantum verifier $V$
for a quantum $k$-prover interactive proof system
such that, for every input $x$ of length $n$,
(i) if $x$ is in $L$,
there exists a set of $m$-message $(q_{\calM}, m q_{\calM})$-restricted quantum provers $P_{1}, \ldots, P_{k}$
without prior entanglement
such that $(P_{1}, \ldots, P_{k}, V)$ accepts $x$ with probability at least $a(n)$,
and (ii) if $x$ is not in $L$,
for all sets of $m$-message $(q_{\calM}, m q_{\calM})$-restricted quantum provers $P'_{1}, \ldots, P'_{k}$ without prior entanglement,
$(P'_{1}, \ldots, P'_{k}, V)$ accepts $x$ with probability at most $b(n)$.

We construct a $km/2$-oracle-call verifier $V^{\QOC}$ of a quantum oracle circuit
as follows.
Let us consider that quantum registers
(collections of qubits upon which various transformations are performed)
$\bfW$,
$\bfM_{i}$, and $\bfP_{i}$, for $1 \leq i \leq k$,
are prepared among the private qubits of the verifier $V^{\QOC}$,
and quantum registers $\bfM$ and $\bfP$
are prepared among the qubits for oracle calls.
$\bfW$ consists of $q_{\calV}$ qubits,
each $\bfM_{i}$ and $\bfM$ consist of $q_{\calM}$ qubits,
and each $\bfP_{i}$ and $\bfP$ consist of $q_{\calP} = m q_{\calM}$ qubits.
Let $\calW^{\QOC}$, each $\calM^{\QOC}_{i}$,
and each $\calP^{\QOC}_{i}$
denote the Hilbert spaces corresponding to the registers $\bfW$,
$\bfM_{i}$, and $\bfP_{i}$, respectively.
Take the Hilbert space $\calV^{\QOC}$
corresponding to the qubits private to the verifier $V^{\QOC}$ as
$\calV^{\QOC} =
\calW^{\QOC} \otimes
\calM^{\QOC}_{1} \otimes \cdots \otimes \calM^{\QOC}_{k} \otimes
\calP^{\QOC}_{1} \otimes \cdots \otimes \calP^{\QOC}_{k}$.
Accordingly, the number of private qubits of $V^{\QOC}$ is
$q^{\QOC}_{\calV}
= q_{\calV} + k(q_{\calM} + q_{\calP})
= q_{\calV} + k(m+1)q_{\calM}$.
Let $\calM^{\QOC}$ and $\calP^{\QOC}$
denote the Hilbert spaces corresponding to
the registers $\bfM$ and $\bfP$, respectively.
Take the Hilbert space $\calO^{\QOC}$
corresponding to the qubits for oracle calls as
$\calO^{\QOC} = \calM^{\QOC} \otimes \calP^{\QOC}$.
Accordingly, the number of qubits for oracle calls
is $q^{\QOC}_{\calO} = q_{\calM} + q_{\calP} = (m+1)q_{\calM}$.

Consider each $V_{j}$, the $j$th quantum circuit of the verifier $V$
of the original quantum $k$-prover interactive proof system,
which acts on
$\calV \otimes \calM_{1} \otimes \cdots \otimes \calM_{k}$.
For each $j$, let $U^{\QOC}_{j}$ be just the same unitary transformation
as $V_{j}$
and $U^{\QOC}_{j}$ acts on
$\calW^{\QOC} \otimes
\calM^{\QOC}_{1} \otimes \cdots \otimes \calM^{\QOC}_{k}$,
corresponding to that $V_{j}$ acts on
$\calV \otimes \calM_{1} \otimes \cdots \otimes \calM_{k}$.
Define the verifier $V^{\QOC}$ of the corresponding quantum oracle circuit
in the following way:
\begin{itemize}
\item
At the first transformation of $V^{\QOC}$,
$V^{\QOC}$ first applies $U^{\QOC}_{1}$,
and then swaps the contents of $\bfM_{1}$
for those of $\bfM$.
\item
At the $((j-1)k+1)$-th transformation of $V^{\QOC}$
for each $2 \leq j \leq m/2$,
$V^{\QOC}$ first swaps the contents of $\bfM$ and $\bfP$
for those of $\bfM_{k}$ and ${\bf P}_{k}$, respectively,
then applies $U^{\QOC}_{j}$,
and finally swaps the contents of $\bfM_{1}$ and $\bfP_{1}$
for those of $\bfM$ and $\bfP$.
\item
At the $((j-1)k+i)$-th transformation of $V^{\QOC}$
for each $2 \leq i \leq k, 1 \leq j \leq m/2$,
$V^{\QOC}$ first swaps the contents of $\bfM$ and $\bfP$
for those of $\bfM_{i-1}$ and $\bfP_{i-1}$, respectively,
then swaps the contents of $\bfM_{i}$ and $\bfP_{i}$
for those of $\bfM$ and $\bfP$.
\end{itemize}

\begin{itemize}
\item[(i)]
In the case the input $x$ of length $n$ is in $L$:\\
In the original $m$-message quantum $k$-prover interactive proof system,
there exist $m$-message $(q_{\calM}, q_{\calP})$-restricted prior-unentangled
quantum provers $P_{1}, \ldots, P_{k}$
that cause $V$ to accept $x$ with probability at least $a(n)$.
Hence, if we let $O_{(j-1)k+i}$
for each $1 \leq i \leq k, 1 \leq j \leq m/2$
be just the same unitary transformation as $P_{i,j}$
($O_{(j-1)k+i}$ acts on
$\calO^{\QOC} = \calM^{\QOC} \otimes \calP^{\QOC}$
corresponding to that $P_{i,j}$
acts on
$\calM_{i} \otimes \calP_{i}$),
it is obvious that
the probability of accepting $x$ by $V^{\QOC}$ with access to $O$
is exactly equal to the one the original $V$ accepts it,
which is at least $a(n)$.
\item[(ii)]
In the case the input $x$ of length $n$ is not in $L$:\\
Suppose that there were an oracle $O'$ that makes the verifier $V^{\QOC}$
accept $x$ with probability more than $b(n)$.
Consider $m$-message $(q_{\calM}, q_{\calP})$-restricted
prior-unentangled provers $P'_{1}, \ldots, P'_{k}$ of the original
$m$-message quantum $k$-prover interactive proof system
such that, for each $1 \leq i \leq k, 1 \leq j \leq m/2$,
$P'_{i,j}$ is just the same transformation as $O'_{(j-1)k+i}$
($P'_{i,j}$ acts on $\calM_{i} \otimes \calP_{i}$
corresponding to that $O'_{(j-1)k+i}$ acts on
$\calM^{\QOC} \otimes \calP^{\QOC}$).
By their construction, it is obvious that
the probability with which these provers $P'_{1}, \ldots, P'_{k}$ can convince the verifier $V$
is exactly equal to the one with which the oracle $O'$ can, which is more than $b(n)$.
This contradicts the assumption.
\end{itemize}
\vspace{-4mm}
\end{proof}

The inclusion $\QMIP^{\mathrm{(n.e.)}} \subseteq \QOC$
immediately follows from Lemma~\ref{Lemma: QOC simulation of QMIP(n.e.)}.
Thus we have the following theorem.

\begin{theorem}
$\QMIP^{\mathrm{(n.e.)}} = \QOC = \NEXP$.
\label{Theorem: QMIP(n.e.) = QOC = NEXP}
\end{theorem}


\section{Conclusions and Open Problems}
\label{Section: Conclusions}

This paper analyzed 
the power of quantum multi-prover interactive proof systems
and gave the $\NEXP$ upper bound for them
in the cases that provers share at most polynomially many prior-entangled qubits.
In particular,
if provers do not share any prior entanglement with each other,
the class of languages
having quantum multi-prover interactive proof systems
was shown equal to $\NEXP$.
Related to these,
if a prover does not have his private qubits,
the class of languages
having quantum single-prover interactive proof systems
was also shown equal to $\NEXP$.

A number of interesting problems remain open
regarding quantum interactive proof systems.

\begin{itemize}
\item
We know very little about the power of
general quantum multi-prover interactive proof systems
with provers sharing arbitrarily many prior-entangled qubits.
Can exponentially many prior-entangled qubits among provers
help a quantum verifier to verify a language not in $\NEXP$?
Does $\NEXP$ have quantum multi-prover interactive proof systems
with prior-entangled provers?
\item
Probabilistic oracle machines are closely related to the theory
of probabilistic checkable proofs~\cite{AroSaf98JACM, AroLunMotSudSze98JACM}.
How is the relation between the quantum oracle circuits
introduced in this paper
and possible quantum analogues of
probabilistic checkable proofs?
\item
In the classical setting the power of one-message multi-prover
interactive proof systems obviously remains same as
that of one-message single-prover ones.
However, as Kobayashi, Matsumoto, and Yamakami~\cite{KobMatYam03quant-ph}
noticed, it might not be so in the quantum setting.
How is the power of one-message quantum multi-prover
interactive proof systems
(both in the cases with and without prior entanglement)?
\end{itemize}


\section*{Acknowledgements}

The authors are grateful to Richard~E. Cleve
for explaining how an entangled pair of provers can cheat
a classical verifier in some cases,
and Lance~J. Fortnow for his valuable comments on writing this paper.
The authors would also like to thank Hiroshi Imai
for his comments and support.




\appendix

\section*{Appendix}

\section{Proof of Theorem~\ref{Theorem: NEXP subseteq QMIP(n.e.)}}
\label{Appendix: NEXP subseteq QMIP(n.e.)}

It is known that every language in $\NEXP$ has
a (classical) multi-prover interactive proof system,
in particular,
a one-round two-prover classical interactive proof system
with exponentially small one-sided error~\cite{FeiLov92STOC}.
Under the assumption that provers do not share any prior entanglement
with each other,
it is easy to show that
a quantum verifier
can successfully simulate such a classical one-round two-prover protocol
(cf.~\cite{Cle00WQCI}).

\begin{proofof}{Theorem~\ref{Theorem: NEXP subseteq QMIP(n.e.)}}
Given a classical $k$-prover interactive protocol,
consider such a quantum $k$-prover protocol without prior entanglement
that a quantum verifier performs measurements in $\{ \ket{0}, \ket{1} \}$ basis
on every qubit of his part
at every time he sends questions to quantum provers and
at every time he receives responses from them,
and for the rest part of computation the quantum verifier behaves in the same manner
as the classical verifier does.
Such a protocol can be simulated without intermediate measurements
by only using unitary transformations~\cite{AhaKitNis98STOC,Gru99Book}.
Furthermore, since there is no prior entanglement
among private qubits of the quantum provers,
such a quantum protocol makes no difference
from a classical protocol
in which a classical verifier chooses a set of $k$ classical provers
probabilistically at the beginning of the protocol.
Therefore, in such a quantum $k$-prover protocol, for every input,
the quantum provers can be only as powerful as the classical provers,
i.e.,
the quantum provers can behave just in the same way as the classical provers do,
while no set of $k$ quantum provers can convince the quantum verifier
with probability more than
the maximum probability with which
a set of $k$ classical provers can convince the classical verifier.

Now we explain in more detail.
Let $L$ be a language in $\NEXP$,
then $L$ has
a one-round two-prover interactive proof system.
Let $V$ be the classical verifier of this one-round two-prover interactive proof system.
We construct a two-message quantum two-prover interactive proof system
by just simulating this classical protocol.

Assume that, just after the classical verifier $V$
has sent questions to the provers $P_{1}$ and $P_{2}$,
the contents of $V$'s private tape, the question to $P_{1}$,
and the question to $P_{2}$ are $v$, $q_{1}$, and $q_{2}$, respectively, 
with probability $p(v, q_{1}, q_{2})$.
Our two-message quantum verifier $V^{\mathrm{(Q)}}$
prepares the quantum registers
$\bfV$, $\bfQ_{1}$, $\bfQ_{2}$, $\bfA_{1}$, and $\bfA_{2}$
among his private qubits.
$V^{\mathrm{(Q)}}$ first stores $v$, $q_{1}$, and $q_{2}$
in $\bfV$, $\bfQ_{1}$, and $\bfQ_{2}$, respectively,
then copies the contents of each $\bfQ_{i}$
to the message qubits shared with a quantum prover $P_{i}^{\mathrm{(Q)}}$.
That is, $V^{\mathrm{(Q)}}$ prepares the superposition
\[
\sum_{v, q_{1}, q_{2}} 
\Bigl(
\sqrt{p(v, q_{1}, q_{2})}
\underbrace{\ket{v}}_{\bfV}
\underbrace{\ket{q_{1}}}_{\bfQ_{1}}
\underbrace{\ket{q_{2}}}_{\bfQ_{2}}
\underbrace{\ket{0}}_{\bfA_{1}}
\underbrace{\ket{0}}_{\bfA_{2}}
\underbrace{\ket{q_{1}}}_{\bfM_{1}}
\underbrace{\ket{0}}_{\bfP_{1}}
\underbrace{\ket{q_{2}}}_{\bfM_{2}}
\underbrace{\ket{0}}_{\bfP_{2}}
\Bigr),
\]
where, for each $i = 1, 2$, $\bfM_{i}$ denotes the quantum register
that consists of the message qubits between $V^{\mathrm{(Q)}}$ and $P_{i}^{\mathrm{(Q)}}$,
and $\bfP_{i}$ denotes the quantum register
that consists of $P_{i}^{\mathrm{(Q)}}$'s private qubits.

Next the quantum provers $P_{1}^{\mathrm{(Q)}}$ and $P_{2}^{\mathrm{(Q)}}$
apply some unitary transformations on their qubits.
Now the state becomes
\begin{eqnarray*}
\lefteqn{
\sum_{v, q_{1}, q_{2}}
\biggl\{
\sqrt{p(v, q_{1}, q_{2})}
\underbrace{\ket{v}}_{\bfV}
\underbrace{\ket{q_{1}}}_{\bfQ_{1}}
\underbrace{\ket{q_{2}}}_{\bfQ_{2}}
\underbrace{\ket{0}}_{\bfA_{1}}
\underbrace{\ket{0}}_{\bfA_{2}}
}\\
& &
\otimes\;
\Bigl(
\sum_{a_{1}}\alpha_{1}(q_{1},a_{1})
\underbrace{\ket{a_{1}}}_{\bfM_{1}}
\underbrace{\ket{\psi_{1}(q_{1},a_{1})}}_{\bfP_{1}}
\Bigr)
\otimes
\Bigl(
\sum_{a_{2}}\alpha_{2}(q_{2},a_{2})
\underbrace{\ket{a_{2}}}_{\bfM_{2}}
\underbrace{\ket{\psi_{2}(q_{2},a_{2})}}_{\bfP_{2}}
\Bigr)
\biggr\}\\
& = &
\sum_{v, q_{1}, q_{2}, a_{1}, a_{2}}
\Bigl(
\sqrt{p(v, q_{1}, q_{2})} \alpha_{1}(q_{1},a_{1}) \alpha_{2}(q_{2},a_{2})\\
& &
\quad\quad\quad
\times
\underbrace{\ket{v}}_{\bfV}
\underbrace{\ket{q_{1}}}_{\bfQ_{1}}
\underbrace{\ket{q_{2}}}_{\bfQ_{2}}
\underbrace{\ket{0}}_{\bfA_{1}}
\underbrace{\ket{0}}_{\bfA_{2}}
\underbrace{\ket{a_{1}}}_{\bfM_{1}}
\underbrace{\ket{\psi_{1}(q_{1},a_{1})}}_{\bfP_{1}}
\underbrace{\ket{a_{2}}}_{\bfM_{2}}
\underbrace{\ket{\psi_{2}(q_{2},a_{2})}}_{\bfP_{2}}
\Bigr),
\end{eqnarray*}
where each
$\alpha_{i}(q_{i},a_{i})$
denotes the transition amplitude and
each $\ket{\psi_{i}(q_{i},a_{i})}$
is a unit vector in the private space of $P_{i}^{\mathrm{(Q)}}$.

Finally,
$V^{\mathrm{(Q)}}$ copies the contents of the message qubits
shared with the quantum prover $P_{i}^{\mathrm{(Q)}}$
to $\bfA_{i}$ to have the following state
\[
\sum_{v, q_{1}, q_{2}, a_{1}, a_{2}}
\Bigl(
\sqrt{p(v, q_{1}, q_{2})} \alpha_{1}(q_{1},a_{1}) \alpha_{2}(q_{2},a_{2})
\underbrace{\ket{v}}_{\bfV}
\underbrace{\ket{q_{1}}}_{\bfQ_{1}}
\underbrace{\ket{q_{2}}}_{\bfQ_{2}}
\underbrace{\ket{a_{1}}}_{\bfA_{1}}
\underbrace{\ket{a_{2}}}_{\bfA_{2}}
\underbrace{\ket{a_{1}}}_{\bfM_{1}}
\underbrace{\ket{\psi_{1}(q_{1},a_{1})}}_{\bfP_{1}}
\underbrace{\ket{a_{2}}}_{\bfM_{2}}
\underbrace{\ket{\psi_{2}(q_{2},a_{2})}}_{\bfP_{2}}
\Bigr),
\]
and does just the same computation as the classical verifier $V$
using $\bfV$, $\bfM_{1}$ and $\bfM_{2}$.
$V^{\mathrm{(Q)}}$ accepts the input if and only if $V$ accepts it.
\begin{itemize}
\item[(i)]
In the case the input $x$ of length $n$ is in $L$:\\
The quantum provers have only to answer
in just the same way as the classical provers do,
and $V^{\mathrm{(Q)}}$ accepts $x$ with probability $1$.
\item[(ii)]
In the case the input $x$ of length $n$ is not in $L$:\\
Since no quantum interference occurs among the computational paths with different
4-tuple $(q_{1}, q_{2}, a_{1}, a_{2})$,
and from the fact that any pair of classical provers cannot convince the classical verifier with probability more than $1/2$ (actually $1/2^{n}$),
it is obvious that,
for any pair of quantum provers,
$V^{\mathrm{(Q)}}$ accepts $x$ with probability at most $1/2$ (actually $1/2^{n}$).
\end{itemize}
\vspace{-4mm}
\end{proofof}


\end{document}